\documentstyle[12pt,epsfig]{article}

\newlength{\dinwidth}
\newlength{\dinmargin}
\newcommand{\bb}{b\bar{b} }
\newcommand{\ra}{\rightarrow }
\newcommand{\GeV}{{\rm GeV} }

\newcommand{\AmS}{{\protect\the\textfont2
  A\kern-.1667em\lower.5ex\hbox{M}\kern-.125emS}}

\begin{document}
\begin{flushright}
IPPP/02/57 \\
DCPT/02/114 \\
3 October 2002 \\
\end{flushright}

\vspace*{2cm}

\begin{center}
{\Large \bf Physics with forward protons at hadron
colliders\footnote{Presented at the 31st International Conference
on High Energy Physics (ICHEP02), Amsterdam, Netherlands, 24--31
July 2002.}}

\vspace*{1cm}
\textsc{V.A. Khoze$^a$, A.D. Martin$^a$ and M.G. Ryskin$^{a,b}$} \\

\vspace*{0.5cm} $^a$ Institute for Particle Physics Phenomenology,
University of Durham, DH1 3LE, UK \\
$^b$ Petersburg Nuclear Physics Institute, Gatchina,
St.~Petersburg, 188300, Russia
\end{center}

\vspace*{1cm}

\begin{abstract}
We emphasize the importance of tagging the outgoing forward
protons to sharpen the predictions for New Physics at the LHC. We
show that exclusive double-diffractive Higgs production, $pp\ra
p+H+p$, followed by the $H\ra\bb$ decay, could play an important
role in identifying a `light' Higgs boson. \vspace{1pc}
\end{abstract}

\section{Introduction}

Recently there has been much interest in studying events with
tagged forward protons at present and forthcoming hadronic
colliders, see, for example,~\cite{INC,DKMOR}. This may not only
allow the luminosity of the colliding protons to be monitored with
high accuracy~\cite{LUM}, but also can provide new ways to
investigate the subtle issues of QCD and to search for the
manifestations of the New Physics. As discussed in~\cite{INC} the
programme with the tagged forward protons is in many aspects
complementary to both the standard physics at hadron colliders and
to studies at a future linear collider.

The physics potential of high energy proton colliders can be
significantly increased by studying exclusive
double-diffractive-like processes of the type $pp\ra p+M+p$. Here
$M$ represents a system of invariant mass $M$, and the $+$ signs
denote the presence of rapidity gaps which separate the system $M$
from the protons. Such processes allow an exceptionally clean
experimental environment to identify New Physics signals (such as
the Higgs boson, SUSY particles, etc., see~\cite{INC}). In such
events we produce a colour-singlet state $M$ which is practically
free from soft secondary particles. Moreover, if forward going
protons are tagged we can reconstruct the `missing' mass $M$ with
good resolution, and so have an ideal means to search for new
resonances and to study threshold behaviour phenomena. We have to
pay a price for ensuring such a clean diffractive signal. In
particular, the diffractive event rate is suppressed by the small
probability, $\widehat{S}^2$, that the rapidity gaps survive soft
rescattering effects between the interacting hadrons, which can
generate secondary particles which populate the gaps, see, for
example,~\cite{KMRsoft,KKMR} and references therein.

\section{Exclusive Higgs Production}

Double-diffractive Higgs production, $pp\ra p+H+p$, at the LHC, is
a good example to illustrate the pros and cons of exclusive
processes. Let us assume a Higgs boson of mass $M_H=120\ \GeV$ and
consider detection in the $\bb$ channel. It is possible to install
proton taggers so that the `missing mass' can be measured to an
accuracy $\Delta M_{\rm missing}\simeq 1\ \GeV$~\cite{DKMOR}. Then
the exclusive process will allow the mass of the Higgs to be
measured in two independent ways. First the tagged protons give
$M_H=M_{\rm missing}$ and second, via the $H\ra\bb$ decay, we have
$M_H=M_{\bb}$, although now the resolution is much poorer with
$\Delta M_{\bb}\simeq10\ \GeV$. The existence of matching peaks,
centered about $M_{\rm missing}=M_{\bb}$, is a unique feature of
the exclusive diffractive Higgs signal. Besides its obvious value
in identifying the Higgs, the mass equality also plays a key role
in reducing background contributions. Another advantage of this
exclusive process, with $H\ra\bb$, is that the leading order
$gg\ra\bb$ background subprocess is suppressed by a $J_z=0$
selection rule~\cite{Liverpool,KMRmm}. The disadvantage is that,
to ensure the survival of the rapidity gaps, the predicted
$H\ra\bb$ cross section is low, $\sigma\simeq2$~fb, corresponding
to a soft survival factor $\widehat{S}^2=0.02$. It is estimated
that there is a factor two uncertainty in this
prediction~\cite{DKMOR}.

For an integrated luminosity of 30~fb$^{-1}$ the number of signal
(background) events for this method of Higgs detection at the LHC
is expected to be 11~(4). These include a factor 0.6 for the
efficiency associated with proton tagging, 0.6 for $b$ and
$\bar{b}$ tagging and 0.5 for the $b,\bar{b}$ jet polar angle cut,
$60^\circ<\theta<120^\circ$ (necessary to reduce the $\bb$ QCD
background)~\cite{DKMOR}.

There exists a huge spread of predictions of the cross sections
for diffractive Higgs production, see, for example,
\cite{CH,PP1,BDPR,SCIH}, which can differ from \cite{INC,DKMOR} by
orders of magnitude. A critical comparison between these
predictions, and an explanation of their differences with our
results, is given in~\cite{Myths}.

A way to check experimentally the reliability of the predictions
of exclusive production is to measure the much larger cross
section for an analogous process: double-diffractive central
production of a pair of high-$E_\perp$ jets~\cite{KMR}. Some of
the existing approaches overshoot the current CDF dijet
data~\cite{CDFjj} by a few orders of magnitude; others are just
normalized to the experimental rates in order to account for the
survival effects. However, the latter procedure is not
unambiguous, since there is no direct way of using the dijet
overshoot factors to correct the expectations for Higgs
production. The perturbative approach~\cite{KMR} predicts about
1~nb~\cite{Liverpool} for the exclusive central production of
dijets with $E_\perp>7\ \GeV$, corresponding to the CDF
kinematics~\cite{CDFjj} to be compared with the observed exclusive
dijet bound of less than 3.7~nb. Moreover, in the case of central
inelastic production (when secondaries are allowed in some central
rapidity interval), our estimates give about 40~nb (with a factor
two uncertainty) for the cross section in the CDF kinematical
range, while the observed value~\cite{CDFjj} is
$43.6\pm4.4\pm21.6$~nb. Bearing in mind that the accuracy of the
theoretical predictions for $E_\perp>7\ \GeV$ jets at the Tevatron
energy is far from perfect, we find this preliminary comparison
quite encouraging.

Another valuable check of the calculations of the soft survival
factor $\widehat{S}^2$ (which originates from the non-perturbative
sector) is the description of diffractive dijet production at the
Tevatron~\cite{CDFfac} in terms of the diffractive structure
functions measured at HERA~\cite{KKMR}. The remarkably good
agreement of these predictions with the CDF measurements is a
confirmation that our calculations of
$\widehat{S}^2$~\cite{KMRsoft,KKMR,KMR} are trustworthy. Moreover,
the new fit to the H1 diffractive data~\cite{HERA} makes the
agreement with the CDF results even better\footnote{Another probe
of the calculations of $\widehat{S}^2$ could come from the
experimental studies of the central $Z$ production by $WW$ fusion,
see, for instance,~\cite{Peter}.}.

Details of the calculation of exclusive Higgs production are given
in Ref.~\cite{INC}. The main sources of the $\bb$ background are,
at leading order, caused by gluon jets being misidentified as a
$\bb$ pair, by a $J_z=2$ admixture due to non-forward protons and
by a $J_z=0$ contribution arising from $m_b\neq 0$. Also there is
a background contribution from $\bb g$ events in which the emitted
gluon is approximately collinear with a $b$ jet. These backgrounds
were considered in detail in Ref.~\cite{DKMOR}, leading to a
prediction of the signal-to-background ratio of about 3. Note that
in \cite{DKMOR} only the $gg\ra\bb g$ hard subprocess was
considered at NLO, and radiation for the spectator, screening
gluon was not discussed. However, this latter process is
numerically small because of the additional suppression of
colour-octet $\bb$ production around $90^\circ$; rotational
invariance around the $b$ quark direction causes the cross section
to be proportional to $\cos^2\theta$ in the $\bb$ c.m. frame.

The cross sections for inclusive and central inelastic diffractive
Higgs production are larger than for exclusive production.
However, for these non-exclusive processes it is hard to suppress
the QCD $\bb$ background and the signal-to-background ratio is
small. Second, we cannot improve significantly the accuracy of the
measurement of the mass of the Higgs boson by tagging the forward
protons and measuring the missing mass.

Recall that, at medium and high luminosity at the LHC, the
recorded events will be plagued by overlap interactions in the
same bunch crossing. For example, at the medium luminosity of
$10^{33}$~cm$^{-2}$s$^{-1}$, an average of 2.3 inelastic events
are expected for each bunch crossing. Hence the rapidity gaps
occurring in one interaction may be populated by particles created
in an accompanying interaction. It is, however, possible to use
detector information to locate the vertices of the individual
interactions and, in principle, to identify hard scattering events
with rapidity gaps. For the exclusive and central inelastic
processes, the use of proton taggers makes it much more reliable
to select the rapidity gap events.

\section{Conclusion}

The Physics menu for LHC studies with tagged forward protons looks
quite attractive and promising~\cite{INC}. In particular, the
exclusive $pp\ra p+H+p$ process has the advantage that the signal
exceeds the background. The favourable signal-to-background ratio
is offset by a low event rate, caused by the necessity to preserve
the rapidity gaps so as to ensure an exclusive signal.
Nevertheless, as shown in~\cite{DKMOR}, the signal for a `light'
Higgs has reasonable significance in comparison to the standard
$H\ra\gamma\gamma$ and $t\bar{t}H$ search modes. Moreover, the
advantage of the matching Higgs peaks, $M_{\rm missing}=M_{\bb}$,
cannot be overemphasized.

We stress that the predicted value of the exclusive cross section
can be checked experimentally. All the ingredients, except for the
NLO correction to the $gg\ra H$ vertex, are the same for our
signal as for exclusive double-diffractive dijet production,
$pp\ra p+ {\rm dijet} + p$, where the dijet system is chosen in
the same kinematic domain as the Higgs boson, that is
$M(jj)\sim120\:{\rm GeV}$. Therefore by observing the larger dijet
production rate, we can confirm, or correct, the estimate of the
exclusive Higgs signal.

\section*{Acknowledgements}

We thank Albert De Roeck, Aliosha Kaidalov and Risto Orava for
valuable discussions, and the EU, PPARC and the Leverhulme Trust
for support.

\end{document}